\documentclass[aps,prl,amsmath,amssymb,superscriptaddress,floatfix,showpacs,twocolumn]{revtex4-1}
\usepackage{bm}
\usepackage{hyperref}
\usepackage[usenames,dvipsnames]{color}

\usepackage{amsmath}

\usepackage{graphicx}
\usepackage{stmaryrd}

\begin{document}

\newcommand{\note}[1]{{\color{red} \bf #1}}
\newcommand{\Proj}[2] { R_{#1}^{} }	
\newcommand{\Projtri}[1] { P_{#1}^{} }	
\newcommand{\ket}[1] {\left| #1 \right\rangle}
\newcommand{\bra}[1] {\left\langle #1 \right|}

\newcommand{\Rchi}{r}
\newcommand{\Lchi}{l}

\newcommand{\swall}[2] {|\eta_{#1}^{#2}\rangle}
\newcommand{\xwall}[2] {|\xi_{#1}^{#2}\rangle}

\newcommand{\swallbra}[2] {\bra{\eta_{#1}^{#2}}}
\newcommand{\xwallbra}[2] {\bra{\xi_{#1}^{#2}}}

\newcommand{\swallbr}[2] { \left\langle \eta_{#1}^{#2} \right. }
\newcommand{\xwallbr}[2] { \left\langle \xi_{#1}^{#2}  \right. }

\newcommand{\xxwall}[2] {\ket{\xi_{#1} \xi_{#2}}}

\newcommand{\xswall}[2] {\ket{\xi_{#1} \eta_{#2}}}

\newcommand{\sswall}[2] {\ket{\eta_{#1} \eta_{#2} }}

\newcommand{\xxwallb}[2] {\ket{\xi_{#1} \xi_{#2} ^* }}

\newcommand{\sswallba}[2] {\ket{\eta_{#1} \eta_{#2} ^{*}}}

\newcommand{\sswallbb}[2] {\ket{\eta_{#1} \eta_{#2} ^{**}}}

\newcommand{\singlet}[2]
{ #1^\uparrow #2^\downarrow-#1^\downarrow #2^\uparrow}

\newcommand{\Hel}[2] 
{ \left\langle  #1 \right| \mathcal{H} \left| #2 \right \rangle }

\newcommand{\xxwallk}[2]
{ \xi \xi \left(#1,#2\right)} 

\newcommand{\xswallk}[2]
{ \xi \eta\left(#1,#2\right)} 

\newcommand{\sswallk}[2]
{ \eta \eta \left(#1,#2\right)}

\newcommand{\footnoteremember}[2]{
\footnote{#2}
\newcounter{#1}
\setcounter{#1}{\value{footnote}}
}
\newcommand{\footnoterecall}[1]{
\footnotemark[\value{#1}]
}

\title{Exact ground states with deconfined gapless excitations for the 3 leg spin--1/2 tube}
\author{Mikl\'os Lajk\'o}
\affiliation{Research Institute for Solid State Physics and Optics, H-1525 Budapest, P.O.B. 49, 
Hungary}
\affiliation{Department of Physics, Budapest University of Technology and Economics, Budafoki \'ut 8, H-1111 Budapest, Hungary}

\author{ Philippe Sindzingre}
\affiliation{  Laboratoire de Physique Th\'eorique de la Mati\`ere Condens\'ee, Universit\'e P. et M. Curie, Paris, France}

\author{Karlo Penc}
\affiliation{Research Institute for Solid State Physics and Optics, H-1525 Budapest, P.O.B. 49, Hungary}

\date{\today}

\pacs{
75.10.Jm, %Quantized spin models, including quantum spin frustration
75.10.Kt	, %Quantum spin liquids, valence bond phases and related phenomena
%75.10.Pq	,  %Spin chain models
75.30.Kz  %Magnetic phase boundaries (including classical and quantum magnetic transitions, metamagnetism, etc.) (for ferroelectric phase transitions, see 77.80.B-; for superconductivity phase diagrams, see 74.25.Dw)
}

\begin{abstract}
We consider a spin-1/2 tube (a three-leg ladder with periodic boundary conditions)  with a Hamiltonian given by two projection operators --- one on the triangles, and the other on the square plaquettes on the side of the tube --- that can be written in terms of Heisenberg and four-spin ring exchange interactions. We identify 3 phases: (i) for strongly antiferromagnetic exchange on the triangles, an exact ground state with a gapped spectrum can be given as an alternation of spin and chirality singlet bonds between nearest triangles; (ii) for ferromagnetic exchange on the triangles, we recover the phase of the spin-3/2 Heisenberg chain; (iii) between these two phases, a gapless incommensurate phase exists. We construct an exact ground state with two deconfined domain walls and a gapless excitation spectrum at the quantum phase transition point between the incommensurate and dimerized phase.
\end{abstract}

\maketitle

%%%%%%%%%%%%%%%%%%%%%%%%%%%%%%%%%%%%%%%%%%%%%%%%%%%%%%%%%%%%%%%%%%%%%%%%%%% 
%    Introduction
%%%%%%%%%%%%%%%%%%%%%%%%%%%%%%%%%%%%%%%%%%%%%%%%%%%%%%%%%%%%%%%%%%%%%%%%%%% 

The projection operator approach to spin models provided significant results on the ground state properties of quantum magnets.
 Examples include the Majumdar--Ghosh Hamiltonian \cite{MG}, a spin--1/2 
 antiferromagnetic Heisenberg chain where the two exact ground state wave functions are given by a product of purely nearest-neighbor valence bonds (pairs of S=1/2 spins forming a singlet) with a gapped excitation spectrum, in accordance with the Lieb--Schultz--Mattis theorem\cite{LSM}.
 The exact  ``valence bond solid'' ground state in the Affleck--Kennedy--Lieb--Tasaki (AKLT) model\cite{AKLT} with gapped excitations is an explicit realization of Haldane's conjecture for $S = 1$ Heisenberg chains\cite{Haldane1983PRL}. 
Further examples include the two--dimensional Shastry--Sutherland model\cite{SSmodel} which has been realized in  SrCu$_2$(BO$_3$)$_2$  \cite{Kageyama}.
In the pyrochlore lattice, Yamashita and Ueda have introduced a model with a macroscopically degenerate ground state \cite{Yamashita2000}. In all these cases the Hamiltonian is a sum of projection operators
\footnote{$P$ is an orthogonal projection if it is self-adjoint and $P = P^2$. This implies that P has only two eigenvalues,  0 for the states in the kernel of P, and 1 for the orthogonal subspace it projects onto.}
and positive semidefinite by construction, so that any state that has 0 energy is an exact ground state.

%%%%%%%%%%%%%%%%%%%%%%%%%%%%%%%%%%%%%%%%%%%%%%%%%%%%%%%%%%%%%%%%%%%%%%%%%%% 
%    Spin tube model
%%%%%%%%%%%%%%%%%%%%%%%%%%%%%%%%%%%%%%%%%%%%%%%%%%%%%%%%%%%%%%%%%%%%%%%%%%% 

\begin{figure}[tb]
\begin{center}
\includegraphics[width=7.0truecm]{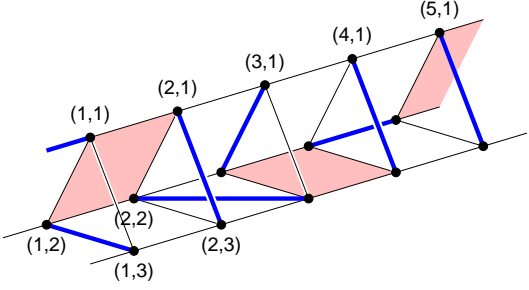}
\caption{The spin tube with a snapshot of valence bond covering from the exact spin--chirality dimerized ground state.  
The shaded plaquettes are not satisfied in this particular configuration, and only quantum resonance with other configurations will make all the plaquettes satisfied. }
\label{fig:tube}
\end{center}
\end{figure}

 Here we extend this approach to a model of spin--1/2's arranged in a 3 leg tube geometry, given by the
\begin{equation}
\label{eq:projHtri}
\begin{split}
\mathcal{H}  = & 
  K_{\triangle}  \sum_{i=1}^L \Projtri{i}
+ K_{\square} \sum_{i=1}^L \sum_{j=1}^3 \Proj{ (i,j)(i+1,j)(i+1,j+1)(i,j+1) }{2}  
\end{split}
\end{equation}
Hamiltonian  (see Fig.~\ref{fig:tube}).  The tube has $L$ triangles and periodic boundary conditions are assumed. 
%(the indices $i$ and $j$  are defined mod L and mod 3, respectively).  
The projector
$\Projtri{i} = (4\mathbf{\tilde S}_i \cdot \mathbf{\tilde S}_i-3)/12 $, where 
$\mathbf{\tilde S}_i=\sum_{j=1}^3 \mathbf{S}_{i,j}$ is the spin operator on the $i^\text{th}$ triangle, gives 1 if the triangle has a total spin of 3/2, and 0 if the spin is 1/2. The projection $\Proj{\alpha}{2}$ acts on the squares that are on the surface of the tube. We denote
$\mathbf{S}_\alpha=\sum_{(i,j)\in \alpha} \mathbf{S}_{i,j}$  as the sum of the spin operators belonging to the $\alpha$ square plaquette, then $\Proj{\alpha}{2}=(\mathbf{S}_\alpha \cdot \mathbf{S}_\alpha) (\mathbf{S}_\alpha \cdot \mathbf{S}_\alpha-2)/24$ projects onto the subspace of states where the total spin of the plaquette $\alpha$ is 2, and gives 0 if the  spin is 0 or 1 (i.e. if a pair of spins on the $\alpha$ square form a valence bond).  
The expanded Hamiltonian using the spin operators reads 
\begin{eqnarray}
\mathcal{H} &=& 
  \sum_{i=1}^L 
    \sum_{j=1}^3 
      \left\{  J_{\perp} \mathbf{S}_{i,j} \cdot \mathbf{S}_{i,j+1} 
+ J_1 \mathbf{S}_{i,j} \cdot \mathbf{S}_{i+1,j}  \right. \nonumber \\
&&{ }+ J_{2} \left(\mathbf{S}_{i,j} \cdot \mathbf{S}_{i+1,j+1} +\mathbf{S}_{i,j} \cdot \mathbf{S}_{i+1,j-1} \right) 
\nonumber\\ 
&&{}  + J_{\text{RE}} \left[ 
( \mathbf{S}_{i,j}\cdot  \mathbf{S}_{i+1,j})( \mathbf{S}_{i,j+1}\cdot  \mathbf{S}_{i+1,j+1}) \right.  \nonumber \\
 &&{}+  ( \mathbf{S}_{i,j}\cdot  \mathbf{S}_{i,j+1})( \mathbf{S}_{i+1,j}\cdot  \mathbf{S}_{i+1,j+1}) \nonumber \\
&&\left.\left.+ ( \mathbf{S}_{i,j}\cdot  \mathbf{S}_{i+1,j+1})( \mathbf{S}_{i,j+1}\cdot  \mathbf{S}_{i,j+1})  \right] \right\}
\label{eq:Hexpand}
\end{eqnarray}
where the intra-triangle $J_\perp= 5 K_\square/6 + 2 K_\triangle /3$, the inter-triangle $J_{1}= 5 K_\square /6$ and $J_2= 5 K_\square/12$, and the ring exchange interaction $J_{\text{RE}} = K_\square/3$. We set $K_{\square} =1$ in the following. 

Spin tubes 
\cite{Cabra1998,Kawano,Orignac,Wang2001,LuscherNoack,PhysRevB.78.054421, Sakai, *Sakai:2010lr, PhysRevB.82.075108,SchmidtandRichter2010} 
are interesting not only since there exist experimental realizations
%, such a [(CuCl$_2$ tachH)$_3$Cl]Cl$_2$ 
\cite{Seeber2004,*Ivanov2010} but also as they are the next step after spin ladders towards two dimension (2D). 
Batista and Trugman, have recently studied  the Hamiltonian that is a sum of the $\Proj{\alpha}{2}$ operators over all the squares of the square lattice and shown that a class of states consisting of nearest neighbor valence bond coverings, where each square plaquette shares a valence bond, have zero energy, i.e. are exact ground states\cite{BatistaTrugman,Nussinov2007}. 
The spin tube is identical to the square latice with period three wrapped in one direction.

%%%%%%%%%%%%%%%%%%%%%%%%%%%%%%%%%%%%%%%%%%%%%%%%%%%%%%%%%%%%%%%%%%%%%%%%%%% 
%    ED
%%%%%%%%%%%%%%%%%%%%%%%%%%%%%%%%%%%%%%%%%%%%%%%%%%%%%%%%%%%%%%%%%%%%%%%%%%% 

\begin{figure}[t]
\begin{center}
\includegraphics[width=0.45 \textwidth]{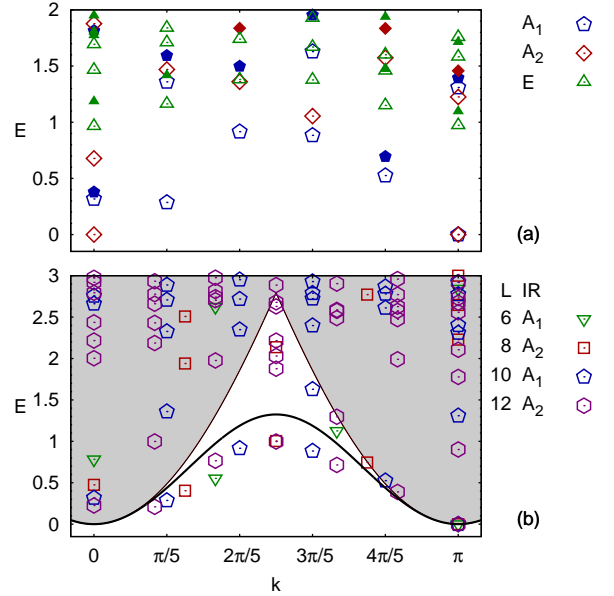}
\caption{(a) Energy spectrum from exact diagonalization of a tube with 10 triangles and $K_\triangle=0$ as a function of momentum along the tube. The empty (filled) symbols denote spin--singlet (triplet) states classified according to the irreducible representations (IR) of $D_3$. (b) Low energy two domain wall excitations in the thermodynamic limit compared to ED spectra of small clusters with symmetry compatible with the variational solution. The thick line below the (shaded) continuum is the bound state. 
}
\label{fig:ed}
\end{center}
\end{figure}

Exact diagonalization (ED) of the Hamiltonian~(\ref{eq:projHtri})  up to $L=12$ triangles (36 sites) showed  for $K_\triangle=0$ an unusual result: for an odd number of triangles it revealed a 0 energy spin--1/2 doublet at $k =0 $ momentum, while for even number we find three singlet ground states: one at $k=0$ and two at $k=\pi$, as shown in Fig.~\ref{fig:ed}(a). The appearance of 0 energy eigenstates  means that they have zero projections with all of the $R_\alpha$ operators in the Hamiltonian. As there is no static covering of valence bonds on the spin--tube where each plaquette is satisfied, this points to a feature that was not encountered in projection Hamiltonians so far. Even more striking is the appearance of the 0 energy ground state for the tubes with odd number of triangles, as in this case necessarily a spin is left unpaired. 
In the following, we will show that this apparent contradiction is resolved due to the quantum mechanical nature of the valence bonds, and that the existence of the 0 energy ground state in the odd size system is related with the fact that $K_\triangle = 0$ is a quantum critical point with gapless deconfined excitations. 

%%%%%%%%%%%%%%%%%%%%%%%%%%%%%%%%%%%%%%%%%%%%%%%%%%%%%%%%%%%%%%%%%%%%%%%%%%% 
%    Effective models
%%%%%%%%%%%%%%%%%%%%%%%%%%%%%%%%%%%%%%%%%%%%%%%%%%%%%%%%%%%%%%%%%%%%%%%%%%% 

Let us begin by considering weakly coupled triangles. The $\Projtri{i}$ splits the 8 states of a triangle into a $\tilde S=3/2$ quadruplet  and two degenerate $\tilde S=1/2$ dublets.
The latter set of four  states constitute the kernel of the $\Projtri{i}$ that can be classified according to the spin $\sigma$ and chirality  $\tau$ degrees of freedom, with an orthonormal basis spanned by the $|\sigma^z_i\tau^z_i\rangle$, where 
\begin{equation}
\label{eq:chi}
\ket{ \sigma_i,\pm 1/2_i}  = | \nu^{\sigma,1}_i \rangle
+ e^{\pm \frac{2\pi i}{3}} | \nu^{\sigma,2}_i \rangle
+ e^{\pm \frac{4\pi i}{3}} | \nu^{\sigma,3}_i \rangle.
\end{equation}
The $|\nu^{\sigma,j}_i\rangle$ are states with a valence bond between sites $j+1,j+2$ and an unpaired spin $\sigma$ on site $j$ of a triangle:
%$|\nu^{\sigma,j}_i\rangle = (|\sigma_{i,j} \uparrow_{i,j+1} \downarrow_{i,j+2}\rangle-|\sigma_{i,j} \downarrow_{i,j+1} \uparrow_{i,j+2}\rangle) /\sqrt{2}$
\begin{equation}
|\nu^{\sigma,j}_i\rangle = \frac{1}{\sqrt{2}}(|\sigma_{i,j} \uparrow_{i,j+1} \downarrow_{i,j+2}\rangle-|\sigma_{i,j} \downarrow_{i,j+1} \uparrow_{i,j+2}\rangle), 
\end{equation}
observing the periodic boundary condition on $j$.
For convenience, we use  $\ket{\Rchi}$ and $\ket{\Lchi}$ for the $\tau^z=\pm 1/2$ chirality.

\begin{figure}
\begin{center}
\includegraphics[width = 7truecm ]{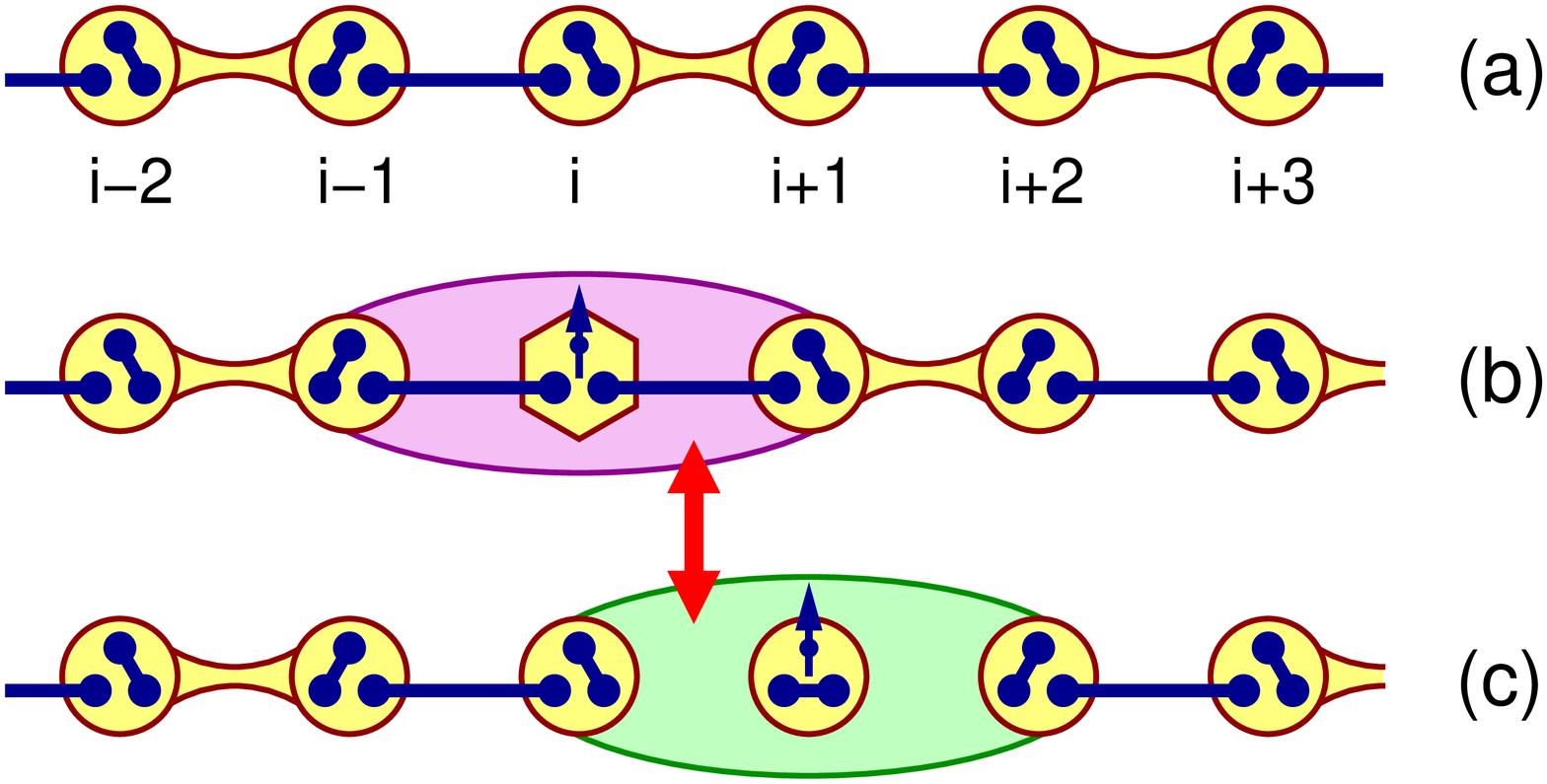}
\caption{(a) Schematic representation of one of the two ground states of $\mathcal{H}$ with alternating spin and chirality valence bonds (a translation by a lattice vector gives the other one).
The small dots inside the circle represent the individual spins of a triangle, thick solid lines denote valence bonds. 
The arcs connecting two circles stand for a chirality singlet bond $\ket{\Rchi\Lchi}-\ket{\Lchi\Rchi}$. 
In (b) and (c) we show the  relevant domain walls in odd length tubes: (b) $\xwall{i}{\uparrow}$, a triangle with $\tilde S=3/2$ (hexagon); (c) the ellipse containing three spin--1/2 triangles with chiralities 
$\ket{ \Rchi\Rchi\Rchi}+\ket{\Lchi\Lchi\Lchi}$ and centered at $i+1$ denotes $\swall{i+1}{\uparrow}$. Both $\xwall{}{\uparrow}$ and $\swall{}{\uparrow}$ have spin--1/2 degrees of freedom. 
}
\label{fig:0or1dwall}
\end{center}
\end{figure}

 In the $ K_\triangle \rightarrow +\infty $  limit the low energy space is spanned by the spin--1/2 states given by Eq.~(\ref{eq:chi}) and the effective spin--chirality  Hamiltonian reads 
\begin{equation}
\mathcal{H}' = 
\frac{5}{9} 
\sum_{i=1}^{L} 
\left(\frac{3}{4} + \bm{\hat\sigma}_i \cdot \bm{\hat \sigma}_{i+1}  \right) 
\left( 1+ \hat\tau_i^+ \hat\tau_{i+1}^- + \hat\tau_i^- \hat\tau_{i+1}^+ \right),
\label{eq:Heff}
\end{equation}
where $\hat \sigma_i$ are the spin--1/2, and $\hat \tau_i^{\pm}$  are the  chirality  (pseudospin--1/2) raising/lowering  operators of the $i^\text{th}$ triangle. This is a particular case of the effective model studied in Refs.~\cite{Kawano,LuscherNoack,Fouet,PhysRevB.78.054421,Sakai, *Sakai:2010lr,PhysRevB.63.064418,*Wang2001,Orignac}, where it has been shown that the spectrum is gapped. 
Since the energy between two neighboring triangles is 0 if either the spins or the chiralities form a singlet, the positive semidefinite $\mathcal{H}'$ has a twofold degenerate exact ground state of alternating spin and chirality singlet bonds \cite{Kolezhuk1998}, shown in Fig. \ref{fig:0or1dwall}(a).

 Actually these two states breaking translational invariance are ground states not only of the effective model (\ref{eq:Heff}), but are exact eigenstates of Eq.~(\ref{eq:projHtri}) for any value of $K_\triangle$, and are ground states for $K_\triangle \geq 0$, when expressed as a linear superpositions of valence bonds\footnoteremember{myfootnote}{See supplementary material after the bibliography for the explicit form of the ground states and the matrix elements of the Hamiltonian between the two domain wall states in the case  $K_\triangle=0$ .}. 
  In this wave function each triangle contains a valence bond -- this makes the $\Projtri{i}$ projections on the triangles happy. The unpaired spins of the triangles form valence bonds that connect pairs of neighboring triangles (they map to the spin--singlet bonds of the effective model), as shown in Fig.~\ref{fig:tube} between triangles 2 and 3, and 4 and 5. The plaquettes between these connected triangles all have a valence bond, so the corresponding $\Proj{\alpha}{}$s give 0. However,
 out of the three plaquettes belonging to a chirality singlet (e.g. triangle 1,2 or 3,4 in Fig.~\ref{fig:tube}) two have valence bond and the third one is seemingly not satisfied. 
  The problem can be resolved if we realize that the three $|\nu^{\sigma,j}_i\rangle$ states used in the definition of  $\ket{ \sigma, \tau }$ in Eq.~(\ref{eq:chi}) are not independent, namely 
$|\nu^{\sigma,1}_{i}\rangle + |\nu^{\sigma,2}_{i}\rangle + |\nu^{\sigma,3}_{i}\rangle = 0$.
So the chirality singlet between the $i^{\text{th}}$ and $i+1^{\text{th}}$ triangles can be expressed as 
$|\nu^{\sigma_{i},1}_i \nu^{\sigma_{i+1},2}_{i+1}\rangle - |\nu^{\sigma_{i},2}_i \nu^{\sigma_{i+1},1}_{i+1}\rangle$. In this expression the plaquettes between the legs $j=1$ and 3, and $j=2$ and 3 both have valence bonds explicitly in each term, thus they are satisfied. Using the linear dependence, this is exactly the same as $|\nu^{\sigma_i,2}_i \nu^{\sigma_{i+1},3}_{i+1}\rangle - |\nu^{\sigma_{i},3}_i \nu^{\sigma_{i+1},2}_{i+1}\rangle$, that makes the third plaquette explicitly satisfied.  Thus we have shown that even though there is no valence bond covering that satisfies all the plaquettes, a linear combination of `imperfect' coverings still constitutes a ground state (i.e. for any plaquette we can choose a basis where a valence bond is explicitly present on that plaquette\cite{NussinovArxiv2006}).

%%%%%%%%%%%%%%%%%%%%%%%%%%%%%%%%%%%%%%%%%%%%%%%%%%%%%%%%%%%%%%%%%%%%%%%%%%% 
%    Odd length tubes
%%%%%%%%%%%%%%%%%%%%%%%%%%%%%%%%%%%%%%%%%%%%%%%%%%%%%%%%%%%%%%%%%%%%%%%%%%% 

For tubes of odd length with periodic boundary conditions the system cannot be covered with alternating spin and chirality singlets, yet ED shows a ground state with zero energy. 
A study of finite size wave functions disclosed that the following two types of domain--wall excitation of the spin--chirality singlet wave function are relevant: one is a $\tilde S=3/2$ triangle connected with valence bonds to the two sides, the other is a $\tilde S=1/2$ triangle connected with chirality triplets to the sides, as shown in Fig.~\ref{fig:0or1dwall}. We use the notation  $\xwall{i}{\sigma}$ and $\swall{i}{\sigma}$ respectively, where $i$ denotes the position of the domain wall. $\xwall{i}{\sigma}$ and $\swall{i}{\sigma}$ have 0 eigenvalue with all the $\Proj{}{}$ in the Hamiltonian, except those that include spins on the $i^{\text{th}}$ triangle. 
Using these states as a variational basis,
we get the following non-vanishing overlaps in Fourier-space: 
$\swallbra{k}{\sigma} \mathcal{H} \swall{k}{\sigma}=5 \left(1- a_L\cos k \right)/6$, 
$\xwallbra{k}{\sigma} \mathcal{H} \xwall{k}{\sigma}=5 \left(1- a_L\cos k \right) / 18$,
and  
$\xwallbra{k}{\sigma} \mathcal{H} \swall{k}{\sigma}=-5 \sqrt{3} \left(\cos k-a_L \right) / 18$, 
while the overlaps are 
$\swallbr{k}{\sigma} \swall{k}{\sigma} = \left(1 - a_L \cos k \right)$ and
$\xwallbr{k}{\sigma} \xwall{k}{\sigma} = 1$, where $a_L=8/2^L$ vanishes for $L \rightarrow \infty$.
It turns out that for $K_\triangle=0$ the 
$\sqrt{3} \xwall{k=0}{\sigma} +  \swall{k=0}{\sigma}$
 is an eigenstate with 0 energy, thus an exact ground state. The propagation of the domain walls in an infinite system is given by the
 \begin{equation}
 \label{1dwallE}
 E^\pm_{1}\left(k \right) =  \frac{5}{36} \left( 4 \pm \sqrt{10+6 \cos{2 k}} \right) 
 \end{equation}
gapless dispersion. This is uncommon for a wave function consisting of short range valence bonds, and is also possibly observed in Ref.~\cite{PhysRevLett.105.067205}. From the Hellmann--Feynman theorem we get that 
$\partial E_{\text{GS}}/\partial K_\triangle = 3 /4$, 
i.e. the energy of the domain wall becomes negative for $K_\triangle<0$, lower than the energy of the dimerized state. This indicates a phase transition at $K_\triangle=0$ between the dimerized and a gapless phase.

%%%%%%%%%%%%%%%%%%%%%%%%%%%%%%%%%%%%%%%%%%%%%%%%%%%%%%%%%%%%%%%%%%%%%%%%%%% 
%    Even length tubes
%%%%%%%%%%%%%%%%%%%%%%%%%%%%%%%%%%%%%%%%%%%%%%%%%%%%%%%%%%%%%%%%%%%%%%%%%%% 
  
\begin{figure}
\begin{center}
\includegraphics[width= 7truecm]{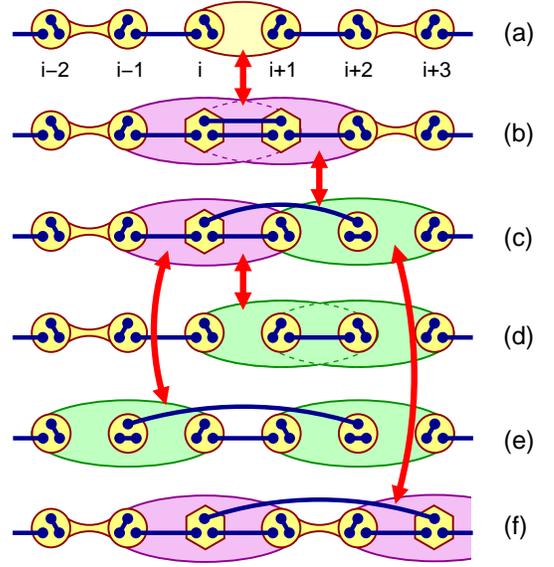}
\caption{Relevant two domain wall configurations in the spin singlet sector for $L$ even.
In (a) the small ellipse denotes a chirality triplet ($\ket{\Rchi\Lchi}+ \ket{\Lchi\Rchi}$), that breaks up into two domain walls.
(c), (e) and (f) are generally given as
$|\xi_{i}^{\uparrow}\eta_{j}^{\downarrow}-\xi_{i}^{\downarrow}\eta_{j}^{\uparrow}\rangle$, 
$|\eta_{i}^{\uparrow}\eta_{j}^{\downarrow}-\eta_{i}^{\downarrow}\eta_{j}^{\uparrow}\rangle$, 
and
$|\xi_{i}^{\uparrow}\xi_{j}^{\downarrow}-\xi_{i}^{\downarrow}\xi_{j}^{\uparrow}\rangle$,
where the domain walls can be arbitrarily separated.  (b) and (d) shows overlapping domain walls, (b) is actually $|\xi_{i}^{\uparrow}\xi_{i+1}^{\downarrow}-\xi_{i}^{\downarrow}\xi_{i+1}^{\uparrow}\rangle$, in (d) the chirality configuration is $\ket{\Lchi\Lchi\Rchi\Rchi} - \ket{\Rchi\Rchi\Lchi\Lchi}$.
Arrows connect states between which the Hamiltonian $\mathcal{H}_{K_\Delta=0}$ has  a nonzero matrix element, the position of the arrows corresponds to the position of $\Proj{}{}$ relevant in the overlap.}
\label{fig:2domainwalls}
\end{center}
\end{figure}

Let us now return to tubes with even number of spins. In this case the relevant excitations are pairs of domain walls that originate from promoting a chirality singlet bond into a chirality triplet, as shown in Fig.~\ref{fig:2domainwalls}. In large systems the domain walls can propagate independently if they are far from each other.  When the domain walls get close they can `overlap' spatially, these components can again be retrieved from results of ED for small systems.
Once all the relevant states  are identified, the overlap matrix and the matrix of the Hamiltonian between these states can be calculated analytically for any finite length systems, allowing us to take the $L\rightarrow \infty$  limit \footnoterecall{myfootnote}.
At $k=\pi$, a particular linear combination, featuring  deconfined domain walls, has 0 energy independent of the system size, and is therefore  the third exact ground state.
The low energy part of the two domain wall spectrum is shown in Fig.~\ref{fig:ed}(b): the two 
independently propagating walls form a continuum, while  the interaction between the walls leads to a  bound state. The bound state is present for any value of $k$, whereas the energy gap between the bound state and the continuum is $\propto k^4$ near $k = 0$ and $\pi$, indicating a fine tuned interaction between the domain walls.
%that can be described by the XXZ--model at exactly the SU(2) symmetric point.  
We note that a finite size gap is closing as $1/L^2$ at $k = 0$ in accordance with the Lieb-Schultz-Mattis theorem, and that the domain walls in the triplet spin sector are also gapless.
 The two domain wall spectrum differs from similar spectra in the Majumdar-Ghosh and AKLT model:
 in the first one a bound state emerges from a gapped two spinon continuum at finite momenta\cite{ShastrySutherland}, while in the latter the gapped bound state is well separated from the continuum\cite{Arovas1988}, even if the interactions are long ranged\cite{JPSJ.63.4565,*PhysRevB.54.9000}

%%%%%%%%%%%%%%%%%%%%%%%%%%%%%%%%%%%%%%%%%%%%%%%%%%%%%%%%%%%%%%%%%%%%%%%%%%% 
%    Phase diagram
%%%%%%%%%%%%%%%%%%%%%%%%%%%%%%%%%%%%%%%%%%%%%%%%%%%%%%%%%%%%%%%%%%%%%%%%%%% 
 
 Next, let us explore what happens for the $K_\triangle<0$, below the point where the gap closes. Using ED, we find a succession of level crossings where the ground state alternates between the $A_1$ and $A_2$ symmetry as we decrease $K_\triangle$ from 0. If we recall that the symmetry of a chirality valence bond is $A_2$, the alternation is related to introducing a pair of domain walls at each level crossing. This phase is gapless and incommensurate, however its precise characterization we leave for a future work. At $K_\triangle \approx -0.23$ the alternation terminates before all the $L$ possible domain walls are added, and the incommensurate phase ends with a first order phase transition.  
  
 For $K_\triangle\alt -0.23$ the model is in the universality class of the spin--3/2 Heisenberg model. In the $K_\triangle \rightarrow  - \infty $ limit, where the spins on the triangles are ferromagnetically coupled, the effective Hamiltonian is the $\tilde S=3/2$ Heisenberg model with bilinear and biquadratic exchanges, that are $7/12$ and $1/18$ in units of $K_\square$, respectively.
% :
%\begin{equation}
%\mathcal{H}'' =\sum_i \left[
%  \frac{45}{32}
%+ \frac{7}{12} \mathbf{\tilde{S}}_i \cdot \mathbf{\tilde{S}}_{i+1}
%+ \frac{1}{18} \left(\mathbf{\tilde{S}}_i \cdot \mathbf{\tilde{S}}_{i+1} \right)^2 \right]  .
%\end{equation}
 EDs on short chains indicate that the low energy spectrum is adiabatically connected to the one of the spin--3/2 Heisenberg model, and is therefore gapless. In the absence of the four-spin ring exchange interaction density matrix renormalization group calculations found a 1$^\text{st}$ order phase transition between the gapped dimerized and the gapless $\tilde S=3/2$ phase \cite{Okunishi,Fouet}, with no signature of the  incommensurate phase.

 \begin{figure}
\begin{center}
\includegraphics[width= 0.4 \textwidth]{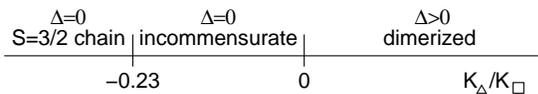}
\caption{Schematic phase diagram of the model.}
\label{fig:schema}
\end{center}
\end{figure}

%%%%%%%%%%%%%%%%%%%%%%%%%%%%%%%%%%%%%%%%%%%%%%%%%%%%%%%%%%%%%%%%%%%%%%%%%%% 
%    Conclusions
%%%%%%%%%%%%%%%%%%%%%%%%%%%%%%%%%%%%%%%%%%%%%%%%%%%%%%%%%%%%%%%%%%%%%%%%%%% 
In conclusion, we have demonstrated that even though the nearest neighbor valence bond coverings can not satisfy all the projection operators in the Hamiltonian, a linear superposition of such coverings is a good ground state. Furthermore, we have constructed an exact ground state at a quantum phase transition point that involves short range valence bonds and deconfined domain walls with a gapless excitation spectrum. Identifying the relevant excitations and combining with exact diagonalization studies, we have conjectured the phase diagram of the model (Fig.~\ref{fig:schema}), with three phases as a function of the interaction strength in the triangles: a gapped dimerized, a gapless incommensurate, and the gapless Luttinger liquid phase of the spin--3/2 Heisenberg chain. 

%
%In conclusion, we have demonstrated that a linear combination of valence bond coverings is an exact ground state, even if the projection Hamiltonian cannot be satisfied by a static valence bond covering on a spin tube. Furthermore, we have constructed an exact ground state at a quantum phase transition point that involves short range valence bonds and deconfined domain walls with a gapless excitation spectrum. Identifying the relevant excitations and combining with exact diagonalization studies, we have conjectured the phase diagram of the model (Fig.~\ref{fig:schema}), with three phases as a function of the interaction strength in the triangles: a gapped dimerized, a gapless incommensurate, and the gapless Luttinger liquid phase of the spin--3/2 Heisenberg chain. 
%
% 
\begin{acknowledgments} 
 We are pleased to acknowledge stimulating discussions with A. Kolezhuk, Z. Nussinov, and S. Miyashita. This work was supported by Hungarian OTKA under Grant No. K73455. Large scale numerical computations were performed at IDRIS (Orsay).
\end{acknowledgments} 

%%%%%%%%%%%%%%%%%%%%%%%%%%%%%%%%%%%%%%%%%%%%%%%%%%%%%%%%%%%%%%%%%%%%%%%%%%% 
%    References
%%%%%%%%%%%%%%%%%%%%%%%%%%%%%%%%%%%%%%%%%%%%%%%%%%%%%%%%%%%%%%%%%%%%%%%%%%% 

\bibliography{artbibV2}{}
\begin{widetext}
\section{Supplementary material}

\subsection{dimerized ground state}

The explicit form of the dimerized wave function of alternating spin and chirality singlets reads as 
\begin{equation}
\sum_{\left\{ \sigma_2, \sigma_4,..., \sigma_{L} \right\}}
\bigotimes \limits_{i=1}^{L/2} \left( -1\right)^ {(1/2-\sigma_{2i}^z)}
\left(|\nu^{-\sigma_{2i-2},1}_{2i-1} \nu^{\sigma_{2i},2}_{2i}\rangle - |\nu^{-\sigma_{2i-2},2}_{2i-1}  \nu^{\sigma_{2i},1}_{2i}\rangle\right) .
\end{equation}
The $|\nu^{-\sigma_{2i-2},1}_{2i-1} \nu^{\sigma_{2i},2}_{2i}\rangle - |\nu^{-\sigma_{2i-2},2}_{2i-1}  \nu^{\sigma_{2i},1}_{2i} \rangle $ term is a chirality singlet between triangles $2i-1$ and  $2i$. 
The spin degrees of freedom of triangles $2i$ and $2i+1$ form a singlet bond $| \uparrow \downarrow \rangle- |\downarrow \uparrow \rangle $, therefore the spin of triangle $2i+1$ is the opposite to that of triangle $2i$, and the $-1$ factor takes care of the antisymmetrization.
\end{widetext}

%$\sum \limits_{\sigma_{2i}} (-1)^{1-\sigma_{2i}}  |\sigma_{2i} \sigma_{2\rangle $

%the spin degree of freedom of  level ${2i-1}$ is the opposite of the spin of $\sigma_{2i-2}$, and the 

%
%\[
%\sum_{\left\{ \sigma_2, \tau_2,\sigma_4,..., \tau_{L} \right\}}
%\bigotimes \limits_{i=1}^{L/2}
%\left( -1\right)^{\sigma_{2i}+\tau_{2i}}
%|\nu^{\sigma_{2i},3/2+\tau_{2i}}_{2i-1} \nu^{\bar{\sigma}_{2i+2},3/2-\tau_{2i}}_{2i}\rangle
%\]
%
%\[
%\sum_{\{\sigma\},\{\tau\}}
%\bigotimes \limits_{i=1}^{L/2}
%\left( -1\right)^{\sigma_{i}+\tau_{i}}
%|\nu^{\sigma_{i},3/2+\tau_{i}}_{2i-1} \nu^{\bar{\sigma}_{i+1},3/2-\tau_{i}}_{2i}\rangle
%\]

\subsection{the ground state with 2--domain walls}
Most of the matrix elements of the Hamiltonian between 2-domain wall states, shown in Fig. 4 in the paper, can be derived from the one domain wall overlaps.
As a reminder, the one domain wall onsite terms
$ \Hel{\xi_m^\sigma}{\xi_m^\sigma} = 5/18$
$ \Hel{\eta_m^\sigma}{\eta_m^\sigma} = 5/6$
and the propagation of a domain wall is described by
$ \Hel{\xi_m ^{\sigma}}{\eta_{m \pm 1 }^{\sigma}} =  - 5\sqrt{3}/36$.
These overlaps are valid for the infinitely long systems ($L\rightarrow\infty$), for finite systems we have other terms scaling as $2^{-L}$.

For tubes of even length, the nonzero overlaps in the thermodynamic limit for $m$ and $n$ are sufficiently separated are:
\begin{eqnarray}
\Hel{\singlet{\xi_m }{\xi_n}}{\singlet{\xi_m }{\xi_n}} &=& 5/9, \nonumber\\
\Hel{\singlet{\eta_m}{\eta_n} }{\singlet{ \eta_m}{\eta_n}} &=& 5/3, \nonumber\\
\Hel{\singlet{\xi_m}{ \eta_n}}{\singlet{\xi_m }{\eta_n} } &=& 10/9, \nonumber\\
\Hel{ \singlet{\xi_m }{\xi_n}}{\singlet{\xi_m }{\eta_{n \pm 1}} } &=& -5 \sqrt{3}/36, \nonumber\\
\Hel{\singlet{\eta_m }{\eta_n}}{\singlet{ \eta_m }{ \xi_{n \pm 1} }} &=&- 5\sqrt{3}/36. \nonumber
\end{eqnarray}
 For $\ket{\singlet{\xi_m }{\eta_n}}$ $m-n$ is even, while for  $ \ket{\singlet{\xi_m }{\xi_n}}$ and $\ket{\singlet{\eta_m }{\eta_n}}$ $m-n$ is odd.
 
 When the two domain walls get close, {\it i.e} they overlap spatially, both the diagonal and off-diagonal matrix element may differ from the general case, and they read
\begin{eqnarray}
\Hel{\singlet{\eta_m }{ \eta_{m+1} }}{ \singlet{\eta_m}{ \eta_{m+1}}}&=&  5/6, \nonumber\\
\Hel{\singlet{\xi_m }{\xi_{m+1}}}{\singlet{\xi_m }{ \xi_{m+1}} }&=& 5/12, \nonumber\\
\Hel{\singlet{\xi_m}{ \eta_{m+2 }}}{\singlet{\xi_m }{\eta_{m+2} } }&=& 5/4, \nonumber\\
\Hel{\singlet{\xi_m}{ \xi_{m+1 }}}{\singlet{\xi_m }{\eta_{m+2} } }&=& -5/18, \nonumber\\
\Hel{\zeta_{m+\frac{1}{2}} }{\zeta_{m+\frac{1}{2}}}&=&  5/6, \nonumber\\ 
\Hel{ \singlet{\xi_m}{ \xi_{m+1}}  } { \zeta_{m+\frac{1}{2} }}&=& - 5 \sqrt{6}/36. \nonumber
\end{eqnarray}
These states are orthonormal ({\it i.e.} the overlap matrix is identity) in the $L\rightarrow\infty$ limit.  The matrix elements given above were used to produce the variational results in Fig 2(b). 

The third exact ground state of $\mathcal{H}_{K_\Delta=0}$ is then given as
\begin{widetext}
\begin{equation}
\Psi_{2dw} =
\sqrt{2} \zeta(\pi) +  2 \sqrt{3}\ket{\xxwallk{\pi}{1}} +
3\sum_ {l=4, \text{odd}}^{L/2}(-1)^{\frac{l-1}{2}}   \ket{\xxwallk{\pi}{l}} -
\sqrt{3} i \sum_ {l=2,\text{even}}^{L-2} (-1)^{\frac{l}{2}} \ket{\xswallk{\pi}{l} } -
\sum_ {l=1, \text{odd}}^{L/2} (-1)^{\frac{l-1}{2}} \ket{ \sswallk{\pi}{ l} }
\end{equation}
This wave function is actually valid for a system of arbitrary length $L$.
Above we use the  Fourier transform with the following phase convention 
\begin{equation}
\begin{split}
\xxwallk{k}{ l} &= \sum_{m} \ket{ \singlet{\xi_m}{\xi_{m, m+l}}} \exp \left(i k \left( m+\frac{l}{2}\right) \right) \\
\sswallk{k}{ l} &= \sum_{m} \ket{ \singlet{\eta_m}{\eta_{m, m+l}}} \exp \left(i k \left( m+\frac{l}{2}\right) \right) \\
\xswallk{k}{ l} &= \sum_{m} \ket{ \singlet{\xi_m}{\xi_{m, m+l}}} \exp \left(i k \left( m+\frac{l }{2}\right) \right) \\
\zeta(k) &=  \sum_{m}  \ket{\zeta_{m+\frac{1}{2} }} \exp \left( i k (m+\frac{1}{2}) \right) 
\end{split}
\end{equation}
for $\xxwallk{k}{l}$ and $\sswallk{k}{l}$  $l$ is odd and  $1\leq l \leq  L/2 $ since $\xxwallk{k}{l} = - \xxwallk{k}{-l}$ ,  while for 
 $\xswallk{k}{l}$  $l$ is even and runs from $2$ to $L-2$.
$m + l/2$ corresponds to the center of mass of the two domain walls, and k is the total wave number.
%check limits and signs
\end{widetext}

\end{document}